\documentclass[twocolumns,aps,prc]{revtex4}

\usepackage{epsfig}
\usepackage{graphicx}

\begin{document}

\title{Search for evidence of two photon contribution in elastic electron proton data.}

\author{E. Tomasi-Gustafsson}
\affiliation{\it DAPNIA/SPhN, CEA/Saclay, 91191 Gif-sur-Yvette Cedex,
France }
\author{G. I. Gakh}
\altaffiliation{Permanent address:
\it NSC Kharkov Physical Technical Institute, 61108 Kharkov, Ukraine}
\affiliation{\it DAPNIA/SPhN, CEA/Saclay, 91191 Gif-sur-Yvette Cedex,
France }

\date{\today}
\pacs{25.30.Bf, 13.40.-f, 13.60.Fz, 13.40.Gp}

\begin{abstract}
We reanalyze the most recent data on elastic electron proton scattering. We look for a deviation from linearity of the Rosenbluth fit to the differential cross section, which would be the signature of the presence of two photon exchange. The two photon contribution is parametrized by a one parameter formula, based on symmetry arguments. The present data do not show evidence for such deviation.
\end{abstract}

\maketitle
\section{Introduction}

Form factors (FFs) characterize the internal structure of composite particles. Since the first measurements  \cite{Ho62}, electromagnetic probes are traditionally preferred to the hadronic beams, because the electromagnetic leptonic interaction is exactly calculable in QED, and one can safely extract the dynamical information about the hadronic vertex. Unfortunately, it is not a direct procedure, because, one has to introduce radiative corrections, which may become very large for the unpolarized cross section at large momentum transfer. Radiative corrections for elastic electron-hadron scattering were firstly calculated by Schwinger \cite{Shwinger} and are important for any discussion of the experimental determination of the differential cross section. They are also calculable in QED, within some assumptions concerning the hadronic interaction. There are standard procedures applied to elastic $eN$ scattering data, but question may arise if the necessary accuracy has been reached \cite{Wa94}. 

The largest value of momentum transfer squared, $Q^2$, where the  electric and magnetic proton FFs $G_{Ep}$ and $G_{Mp}$ have been extracted from the unpolarized cross section by the Rosenbluth method \cite{Ro50} is $Q^2\simeq 8.9$ GeV$^2$ \cite{An94} and further extraction up to $Q^2\simeq 31$ GeV$^2$ \cite{Ar75} assumes  $G_{Ep}=G_{Mp}/\mu_p$ ($\mu_p=2.79$ is the magnetic moment of the proton). In order to show the sensitivity of the measurements with respect to $Q^2$, FFs are often normalized to a dipole function:
\begin{equation}
G_D(Q^2)=(1+Q^2 [\mbox{GeV}^2]/0.71)^{-2}.
\label{eq:dipole}
\end{equation}
Recently, new developments, due to the very precise and surprising data obtained at the Jefferson Laboratory (JLab), in $\vec e+p\to e+\vec p$ elastic scattering \cite{Jo00,Ga02}, based on the polarization transfer method  show that the electric and magnetic distributions in the proton are different. 

The application of the polarization transfer method, proposed about 30 years ago \cite{Re68}, has been possible only recently, as it needs high intensity polarized beams, large solid angle spectrometers and advanced techniques of polarimetry in the GeV range. Experiments have been performed at JLab up to $Q^2=5.6$ GeV$^2$ and an extension up to 9 GeV$^2$ is in preparation \cite{04108}. 

The following parametrization for the ratio $R$ of the electric and magnetic form factors well describes these experimental data \cite{Ga02}:
\begin{equation}
R=\mu_p G_{Ep}/G_{Mp}=1-0.13(Q^2~[\mbox{GeV}^2]-0.04)
\label{eq:brash}
\end{equation}
which implies that the ratio monotonically decreases and deviates from unity, as $Q^2$ increasing, reaching a value of $\simeq$ 0.3 at $Q^2\simeq 5.5 $ GeV$^2$.

Therefore, a clear discrepancy appears between the $Q^2$-dependence of the ratio
$R$ of the electric to the magnetic proton form factors, whether derived with the standard Rosenbluth separation or with the polarization transfer method. This statement is confirmed by a reanalysis of the existing data \cite{Ar04a} and by recent measurements \cite{Ch04,Ar04}. This discrepancy is very puzzling, as no evident experimental bias has been found in the data or in the method used, and it has been sources of different speculations. 

One has to stress, at this point, that the observables are, on one side the differential cross section and on the other side, the polarization transferred to the scattered proton. The discrepancy is not at the level of the observables: FFs extracted with the polarization method, are not incompatible with the measured cross section: it has been shown that, calcualting  $G_{Mp}$ from the measured cross section with the constraint on the ratio $R$ from polarization measurements, leads to a renormalization of $G_{Mp}$ of the order of 2-3\% only, with respect to the Rosenbluth data \cite{Br03}, well inside the error bars. 

Instead, the data  are not consistent, if one refers to the slope of the $\epsilon$ dependence of the reduced cross section, ($\epsilon $ is the polarization of the virtual photon, $\epsilon=[1+2(1+\tau)\tan^2(\theta/2)]^{-1}$,$0\le\epsilon\le 1$ ) which is  directly related to $G_{Ep}$. The difference of such slope especially appears with respect to the last, precise data \cite{Ar04}.

A possible question arises on the validity of the one-photon mechanism at large $Q^2$, and, generally, on the radiative corrections to the differential cross section and to polarization observables in elastic $eN$-scattering. If these corrections are large (in absolute value) for the differential cross section \cite{Mo69},  a simplified estimation of the radiative corrections to polarization phenomena \cite{Ma00} shows that they are small for the ratio $P_L/P_T$ of longitudinal to transverse polarization of the proton emitted in the elastic collision of longitudinally polarized electrons with an unpolarized proton target.

In the standard calculations of the radiative corrections \cite{Mo69}, the two-photon exchange mechanism is only partially taken into account considering the special part of the complicated loop integral, where one virtual photon carries all the momentum transfer and the second virtual photon is almost real. This contribution allows to overcome the problem of the 'infrared' divergence. But it has been pointed out \cite{Gu73}  that, at large momentum transfer, the role of another part of the integral, where the momentum transfer is shared between the two photons, can be relatively increased, due to the steep decreasing of the electromagnetic form factors with $Q^2$. This effect can eventually become so large (especially at large $Q^2$) that the traditional description of the electron-hadron interaction in terms of electromagnetic currents (and electromagnetic form factors) can become incorrect.

Numerous tests of the possible $2\gamma$ contribution for elastic $ep$ scattering have been done in the past, using different methods: test of the linearity of the Rosenbluth formula for the differential cross section, comparison of the $e^+p$ and $e^-p$-cross sections, attempts to measure various T-odd polarization observables, but no effect was visible beyond the precision of the experimental data. Only recently the non-zero T-odd asymmetry in the scattering of transversally polarized electrons, $\vec e^-+p\to e^-+p$ has been detected \cite{We01,Ma05}. 

Following the arguments of  Ref. \cite{Gu73}, the relative contribution of two-photon exchange, compared to the main (i.e., one-photon) contribution should be visible at smaller $Q^2$ for heavier targets: $d$, $^3\!He$,  $^4\!He$, because the corresponding form factors decrease faster in comparison with protons as $Q^2$ increases. One would expect to observe the two-photon contribution in $ed$-scattering at $Q^2\ge 1$ GeV$^2$, and for $eN$-scattering at larger momentum transfer, $Q^2\simeq 10$ GeV$^2$. In Ref. \cite{Re99} the possible effects of $2\gamma$-exchange have been estimated from  the precise data on the structure function $A(Q^2)$, obtained at JLab in electron deuteron elastic scattering, up to $Q^2=6$ GeV$^2$ \cite{Al99,Ab99}. The possibility of a $2\gamma$-contribution has not been excluded by this analysis, starting from $Q^2=1$ GeV$^2$, and the necessity of dedicated experiments was pointed out. 

The complete calculation of the $2\gamma$-contribution to the amplitude of the $e^{\pm} p\to e^{\pm} p$-process requires the knowledge of the matrix element for the double virtual Compton scattering, $\gamma^*+N\to\gamma^*+N$, in a large kinematical region of colliding energy and virtuality of both photons, and can not be done in a model independent form. However, one can partially reconcile the data, assuming  that the radiative corrections contribute to the cross section linearly in $\epsilon$ \cite{Bl03,Gu03}, or in framework of other model-dependent assumptions \cite{Chen04}. These attemps give quantitatively different results, but have an effect in decreasing the slope of the differential cross section as a function of $\epsilon$.   

We previuosly showed \cite{Re99,Re1,Re03t,Re04a}, in a model independent way,  that the presence of two-photon exchange destroys the linearity of the Rosenbluth fit, inducing a specific dependence of the differential cross section on the variable $\epsilon$, which is especially large for $\epsilon \to 1$. 

The purpose of this paper is to re-analyse the data about the differential $ep$ scattering, with respect to the deviation of linearity of the Rosenbluth fit. The two-photon contribution is parametrized according to a simple one parameter form, which satisfies the necessary symmetry properties of the $\epsilon$ dependence. A factorisation is assumed for the $Q^2$ and $\epsilon$ dependent terms.
 
\section{Parametrization of the $2\gamma$ contribution}

The traditional way to measure electromagnetic proton form factors consists in the measurement of the $\epsilon$ dependence of the reduced elastic differential cross section $\sigma_{red}$, at fixed $Q^2$. Assuming that the interaction occurs through the exchange of one photon, one can write:
\begin{equation}
\sigma_{red}(Q^2,\epsilon)=\epsilon(1+\tau)[1+(2E/m)\sin^2(\theta/2)]
\displaystyle\frac
{1}{\sigma_{Mott}}\displaystyle\frac{d\sigma}{d\Omega}=\tau G_{Mp}^2(Q^2)+\epsilon G_{Ep}^2(Q^2),
\label{eq:sred}
\end{equation}
with $\tau=Q^2/(4m^2)$, $m$ is the proton mass, $E$ and $\theta$ are the incident electron energy and the scattering angle of the outgoing electron, respectively.

The Rosenbluth separation \cite{Ro50} allows to extract the electric and the magnetic form factors, from a linear fit to the reduced cross section, as a function of $\epsilon$, at a fixed $Q^2$. The slope is directly related to $G_{Ep}$ and the intercept to $G_{Mp}$. However, due to the coefficient $\tau$, the weight of the magnetic contribution becomes larger, as the momentum transfer increases, reducing the sensitivity of the elastic cross section to the electric contribution. As an example, at $Q^2\simeq 4$ GeV$^2$, the term related to the electric FF contributes for less than 10\% to the reduced cross section, assuming the dipole scaling, whereas it would be as low as 2\%, if one assumes the $Q^2$ dependence of $R$ from Eq. (\ref{eq:brash}). One can estimate the level of precision required to extract the electric FFs from the experimental cross section at large $Q^2$. 

In presence of $2\gamma$ exchange, Eq. (\ref{eq:sred}) can be rewritten in the following general form:
\begin{equation}
\sigma_{red}(Q^2,\epsilon)= 
\epsilon G_E^2(Q^2)+\tau G_M^2(Q^2) +
\alpha F(Q^2,\epsilon), 
\label{eq:sred2}
\end{equation}
where $\alpha=e^2/(4\pi)$, and $F(Q^2,\epsilon)$ is a real function (of both independent variables $Q^2$ and $\epsilon$), describing the effects of the $1\gamma\bigotimes 2\gamma$ interference.

Note that, in the general case, $F(Q^2,\epsilon)$ contains two different contributions, one due to the effect of the two-photon contribution on the form factors $G_{E,M}$, and another one due to the third spin structure in the matrix element of $eN$ scattering, induced by $2\gamma$ exchange. Both contributions  can be calculated only in framework of some model, considering different intermediate states for the $2\gamma$ box diagrams.

In such situation, any model independent statement concerning the function $F(Q^2,\epsilon)$ is important, in particular concerning its $\epsilon$ dependence, due to the large sensitivity of the extraction of $G_E^2$ to the additional contribution $F(Q^2,\epsilon)$, at large $Q^2$.

We proved earlier \cite{Re1}, that the C-invariance and the crossing symmetry of hadron electromagnetic interaction, result in the following symmetry properties of $F(Q^2,\epsilon)$ with respect to the variable $x=\sqrt{\displaystyle\frac{1+\epsilon}{1-\epsilon}}$:
\begin{equation}
F(Q^2,x)=-F(Q^2,-x).
\label{eq:sf}
\end{equation}
This means that the $1\gamma\bigotimes2\gamma$ contribution $F(Q^2,\epsilon)$ has an essential non linear $\epsilon$ dependence, at any $Q^2$. For example, the additional (third) spin structure is generating the following $\epsilon$-dependence:
\begin{equation}
F(Q^2,x)\to \epsilon \sqrt{\displaystyle\frac{1+\epsilon}{1-\epsilon}}f^{(T)}(Q^2,\epsilon )~\mbox{or}~
F(Q^2,x)\to  \sqrt{\displaystyle\frac{1+\epsilon}{1-\epsilon}}f^{(A)}(Q^2,\epsilon ),
\label{eq:sfa}
\end{equation}
where the upper index, $(T)$ or $(A)$, corresponds to the tensor \cite{Re03t} or axial \cite{Re04a} parametrization for the third amplitude.

In order to estimate the upper limit for a possible $2\gamma$ contribution to the differential cross section and the corresponding changing to $G_{E,M}(Q^2)$, we analyzed four sets of data \cite{Wa94,An94,Ch04,Ar04}, applying Eq. (\ref{eq:sred2}) with the following parametrization for $F(Q^2,\epsilon)$:
\begin{equation}
F(Q^2,\epsilon)\to \epsilon 
\sqrt{\displaystyle\frac{1+\epsilon}{1-\epsilon}}f^{(T)}(Q^2).
\label{eq:sfit}
\end{equation} 
It is important to stress that Eq. (\ref{eq:sfit}) is a simple expression which contains the necessary symmetry properties of the $1\gamma\bigotimes 2\gamma$ interference, through a specific (and non linear) $\epsilon$ dependence.

We checked that the parametrization:
$$
F(Q^2,x)\to  \sqrt{\displaystyle\frac{1+\epsilon}{1-\epsilon}}f^{(A)}(Q^2)
$$
gives qualitatively similar results.

For the $Q^2$ dependence  of $f^{(T,A)}$ we take:
\begin{equation}
f^{(T,A)}(Q^2)=\displaystyle\frac{C}{(1+ Q^2\mbox{[GeV] }^2/0.71)^2(1+ Q^2\mbox{[GeV]}^2/m^2_{T,A})^2},
\label{eq:sfq}
\end{equation}
where $C$ is a fitting parameter, $m_{T,A}$ is the mass of a tensor or an axial meson with positive C-parity. Taking $m_{T,A}^2\simeq 1.5$ GeV$^2$ (typical value of the corresponding mass), one can predict that the relative role of the $2\gamma$ contribution should increase with $Q^2$. 

\section{Results and Discussion}

In presence of $2\gamma$, the dependence of the reduced cross section on $\epsilon$ can be parametrized as a function of three parameters, $G_E^2$, $G_M^2$ and $C$, according to Eqs. (\ref{eq:sred2}) and (\ref{eq:sfq}).

\begin{figure}
\begin{center}
\includegraphics[width=17cm]{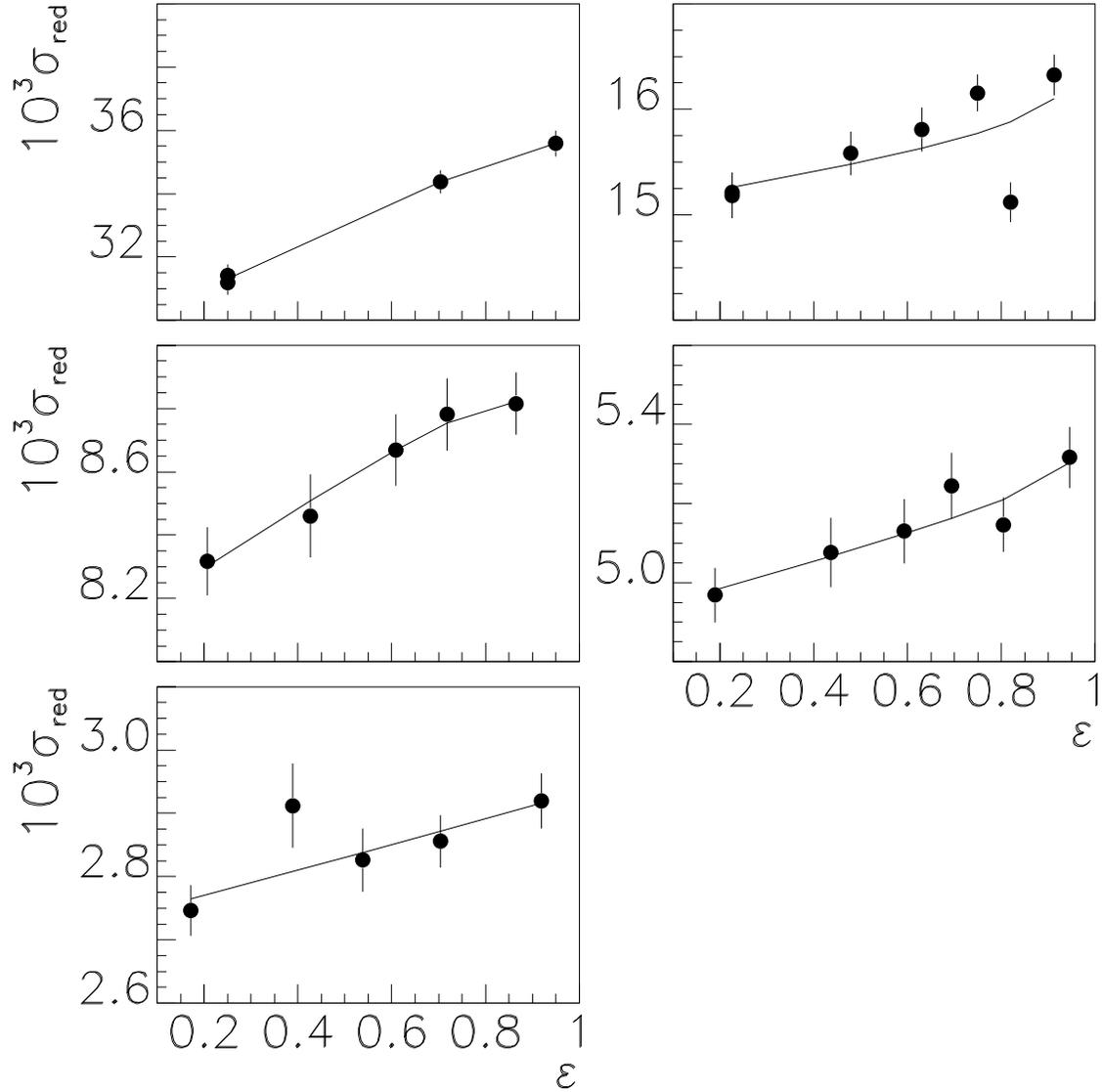}
\caption{\label{fig:fig1} Reduced cross section  at $Q^2$=1.75, 2.5, 3.25, 4, and 5  GeV$^2$. Data are from Ref. \protect\cite{An94}. The lines are three-parameter fits according to Eqs. (\ref{eq:sred2}) and (\ref{eq:sfq}).}
\end{center}
\end{figure}

In Fig. \ref{fig:fig1}, we show $\sigma_{red}$ as a function of the variable $\epsilon$ for different $Q^2$ values,  together with the results of the three parameters fit,  for the data from Ref. \cite{An94}. Fits of similar quality can be obtained for all the considered sets of data.

\begin{figure}
\begin{center}
\includegraphics[width=17cm]{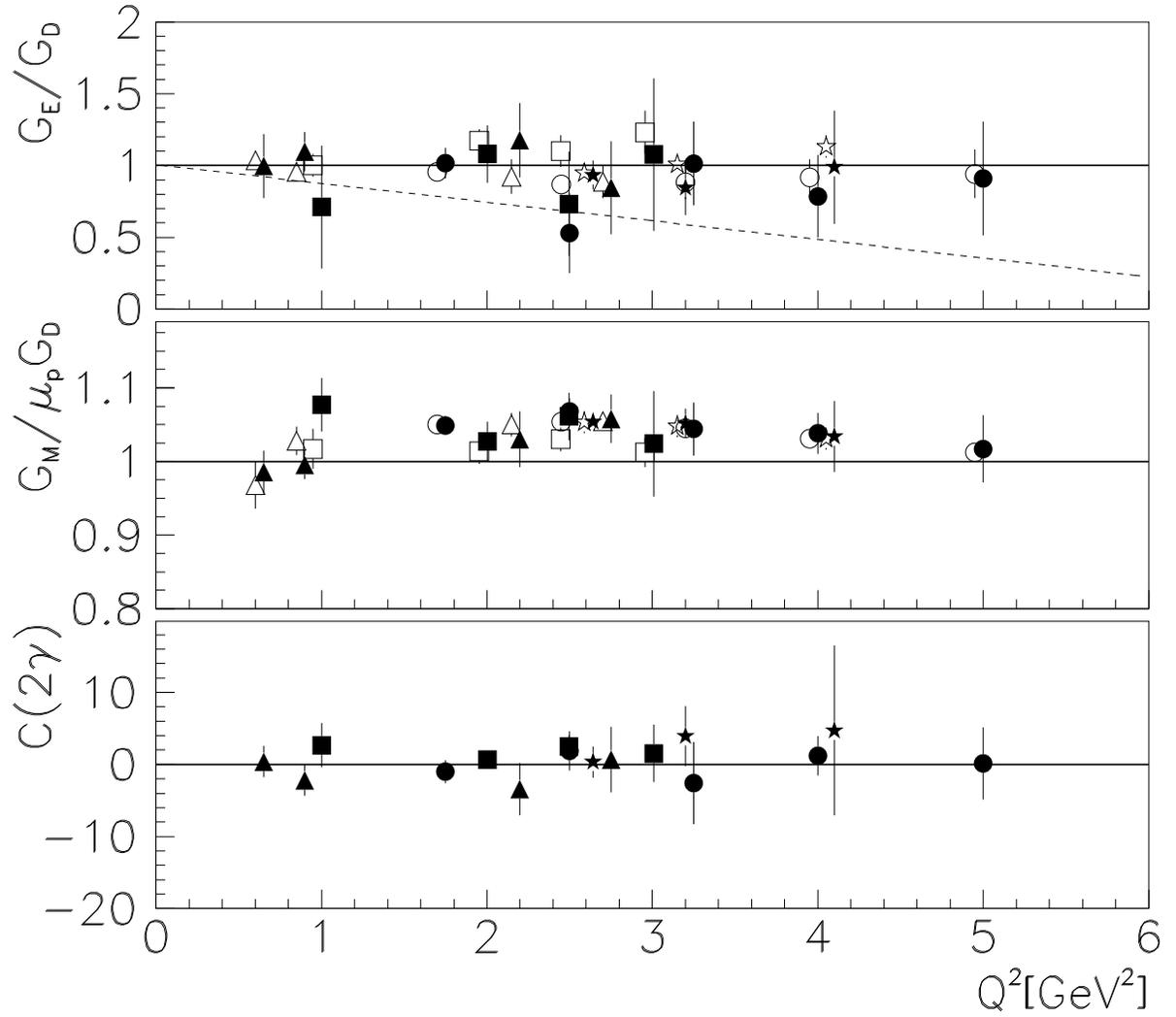}
\caption{\label{fig:fig2} From top to bottom: $G_{Ep}/G_D$, $G_{Mp}/\mu_p G_D$, and two-photon contribution, $C$ according to Eq. (\protect\ref{eq:sfq}) with $m_{T,A}^2=1.5$ GeV$^2$. Data are from Refs. \protect\cite{Wa94} (squares); \protect\cite{An94} (circles); \protect\cite{Ch04} (triangles); \protect\cite{Ar04} (stars). The present results (including $2\gamma$ contribution) are shown as solid symbols. The abscissa of the published data (open symbols) is shifted by -100 KeV, for clarity. The expectation from Eq. (\ref{eq:brash}) is shown as a dashed line.}
\end{center}
\end{figure}

In Fig. \ref{fig:fig2}, from top to bottom, the electric and magnetic FFs, as well as the two photon parameter $C$, are shown as a function of $Q^2$ (solid symbols). The previously published data, derived from the traditional Rosenbluth fit are also shown (open symbols). The expectation from Eq. (\ref{eq:brash}) is shown as a dashed line in Fig. \ref{fig:fig2}(top).

The numerical values of FFs and of the $2\gamma$-coefficient for the data set analyzed here are reported in Table \ref{tab:1}. The resulting parameter $C$ is compatible with zero. The effect of including a third fitting parameter, the $2\gamma$ term, is to increase the error on FFs.  For the  three  points of Ref. \cite{Ch04}, at higher $Q^2$, the coefficient $C$ becomes quite large, but is still compatible with zero within the error. For this reason, these points are reported in the table but not in Fig. \ref{fig:fig2}.
\begin{table}
\begin{tabular}{|c|c|c|c|c|c|c|}
\hline\noalign{\smallskip}
$Q^2$[GeV$^2$] & 
$\displaystyle\frac{G_E}{G_D}\pm\Delta\left ( \displaystyle\frac{G_E}{G_D}\right ) $ & $\displaystyle\frac{G_M}{\mu_p G_D}\pm\Delta\left ( \displaystyle\frac{G_M}{\mu_p G_D}\right )$ & $C \pm\Delta C$ & $\chi^2/n$&$N_{ points}$& Ref.\\
\noalign{\smallskip}\hline\noalign{\smallskip}
1 &0.7 $\pm$  0.4 & 1.08 $\pm$  0.04 &3 $\pm$ 3 & - &3&\protect\cite{Wa94} \\
2 &1.1 $\pm$  0.2 & 1.03 $\pm$  0.03 &1 $\pm$ 1 &0.1 &8&\protect\cite{Wa94} \\
2.5 &0.7 $\pm$  0.4 & 1.06 $\pm$ 0.03 &2 $\pm$ 1 &0.1 &6&\protect\cite{Wa94} \\
3 &1.1   $\pm$  0.5 & 1.02 $\pm$ 0.07 &1 $\pm$ 4 &0.1 &5&\protect\cite{Wa94} \\
\hline\hline
1.75& 1.0 $\pm$ 0.1& 1.05 $\pm$ 0.01 &-1 $\pm$ 2&0.1 &4&\protect\cite{An94} \\
2.5 & 0.5 $\pm$ 0.3& 1.07 $\pm$ 0.02 &2  $\pm$ 3&0.8 &7&\protect\cite{An94} \\
3.25& 1.0 $\pm$ 0.3& 1.04 $\pm$ 0.03 &-3 $\pm$ 6& 0.1 &5&\protect\cite{An94} \\
4.0 & 0.8 $\pm$ 0.3& 1.04 $\pm$ 0.03 &1  $\pm$ 3&0.1 &6&\protect\cite{An94} \\
5.0 & 0.9 $\pm$ 0.4& 1.02 $\pm$ 0.05 &0  $\pm$ 5&0.3&5&\protect\cite{An94} \\
\hline\hline
0.65& 1.0 $\pm$ 0.2& 0.98  $\pm$   0.03& 0 $\pm$ 2&-&3&\protect\cite{Ch04} \\
0.9& 1.1  $\pm$ 0.1 & 0.99 $\pm$ 0.02 & -2 $\pm$ 2&-&3&\protect\cite{Ch04} \\
2.2& 1.2  $\pm$ 0.3& 1.03  $\pm$ 0.04&  -3 $\pm$ 3&-&3&\protect\cite{Ch04} \\
2.75& 0.8 $\pm$ 0.3& 1.06  $\pm$ 0.03&  0. $\pm$ 4&-&3&\protect\cite{Ch04} \\
3.75& 1.6 $\pm$ 0.5& 1.0  $\pm$ 0.1&  -17 $\pm$ 15&-&3&\protect\cite{Ch04} \\
4.25& 1.8 $\pm$ 0.6& 1.0  $\pm$ 0.2&  -19 $\pm$ 26&-&3&\protect\cite{Ch04} \\
5.2& 4  $\pm$ 2& 0.8  $\pm$ 1.5&  -166 $\pm$ 195&-&3&\protect\cite{Ch04} \\
\hline\hline
2.64& 0.9 $\pm$ 0.1& 1.05 $\pm$ 0.01& 0 $\pm$ 2&0.1&5&\protect\cite{Ar04} \\
3.2&  0.8 $\pm$ 0.2& 1.05 $\pm$ 0.02& 4 $\pm$ 4&0.5&4&\protect\cite{Ar04} \\
4.1&  0.9 $\pm$ 0.4& 1.03 $\pm$ 0.05& 5 $\pm$ 12&-&3&\protect\cite{Ar04} \\
\noalign{\smallskip}\hline
\end{tabular}
\caption{Form factors and $2\gamma$-coefficient, Eq. (\protect\ref{eq:sfq}).}
\label{tab:1}       
\end{table}
The present analysis does not give any evidence of a $2\gamma$ contribution, as the possible $2\gamma$ term always vanishes, within the error bars. The new values of $G_E$ and $G_M$ are compatible with the published values, deduced in frame of a standard $\epsilon$ fit. The $\chi^2/n$ (where $n$ is the number of degrees of freedom, i.e., the number of points $N$ minus the number of parameters which is equal to 3) is always smaller than unity. Varying $m^2_{T,A}$ from 1 to 3 GeV$^2$ slightly modifies the coefficient $C$ and its error, being always compatible with zero. In Fig. \ref{fig:fig2} (bottom) no systematic dependence of $C$ with $Q^2$ is visible.  The three points at the highest $Q^2$ from Ref. \cite{Ch04} are not shown in the figure due to their large error bar, which, in our opinion, prevents from any conclusion on a possible systematic trend. 

The average results on this parameter is $<C>=0.5\pm 0.6$. No deviation from zero is seen also for the individual sets of data, within the errors.  Taking into account the $\alpha$ term which is included in Eq. (\ref{eq:sred2}), the deviation of linearity, which is quantified by the parameter $C$ is smaller than 1\%.  We also evaluated the relative weight of the different contributions to the cross section. We calculated the ratios of the magnetic, electric and two photon contributions to the reduced cross section and averaged for all the  considered values of $Q^2$ and $\epsilon$. We find, as expected, that the dominating term is the magnetic term with a contribution at the level of ($95.9 \pm 0.3$)\%. The electric term contributes for ($3.8\pm 0.3$)\% and the average for the two photon term is ($0.05\pm 0.05$)\%.

Furthermore, no visible systematic effect appears, that could allow, at least, to guess the sign of the 2$\gamma$ coefficient. If one could determine at least the sign of the interference contribution, it would be possible to predict the relative value of the differential cross section for $e^-p$ and $e^+p$-scattering. The following relation holds:
\begin{equation}
\displaystyle\frac
{\sigma(e^- p)-\sigma(e^+ p)}{\sigma(e^- p)+\sigma(e^+ p)}=
\displaystyle\frac
{ \alpha F(Q^2,\epsilon)}{ \epsilon G_E^2(Q^2)+\tau G_M^2(Q^2)}.
\label{eq:eqtw}
\end{equation}
Another approach - introducing $ad-hoc$ a linear $\epsilon$ contribution to the differential cross section - results in a corresponding decreasing of the $G_E^2$ contribution, bringing the results in agreement with the polarization measurements. This can not be considered a proof of the presence of the $2\gamma$ contribution, but shows that a small correction to the slope of the reduced cross section is sufficient to solve the discrepancy. Such correction may arise, for example, from a revision of the standard procedure of the calculation of radiative corrections, especially at $\epsilon\to 1$, which would produce a large effect on the value of $G_E^2$, as extracted from the Rosenbluth fit.

The radiative corrections are taken into account in the data analysis, 
according to the prescription of Ref. \cite{Mo69}. Typically, the experimental results on the measured elastic cross section, are corrected by a global factor $\delta_R$: 
\begin{equation}
\sigma^{red}= \sigma^{red}_{meas}e^{\delta_R}\simeq \sigma^{red}_{meas}(1+\delta_R).
\label{eq:sreda}
\end{equation}
This factor {\it contains a large $\epsilon$ dependence}, and a smooth $Q^2$ dependence,  and it is common for the electric and magnetic part. At the largest $Q^2$ considered here, it can reach 30-40\%. A linear $\epsilon$ dependence can be considered a good approximation for the radiative corrections:
$\delta_R=\epsilon \delta '$, but our conclusions hold for any monotonically increasing function. 

\begin{figure}
\begin{center}
\includegraphics[width=17cm]{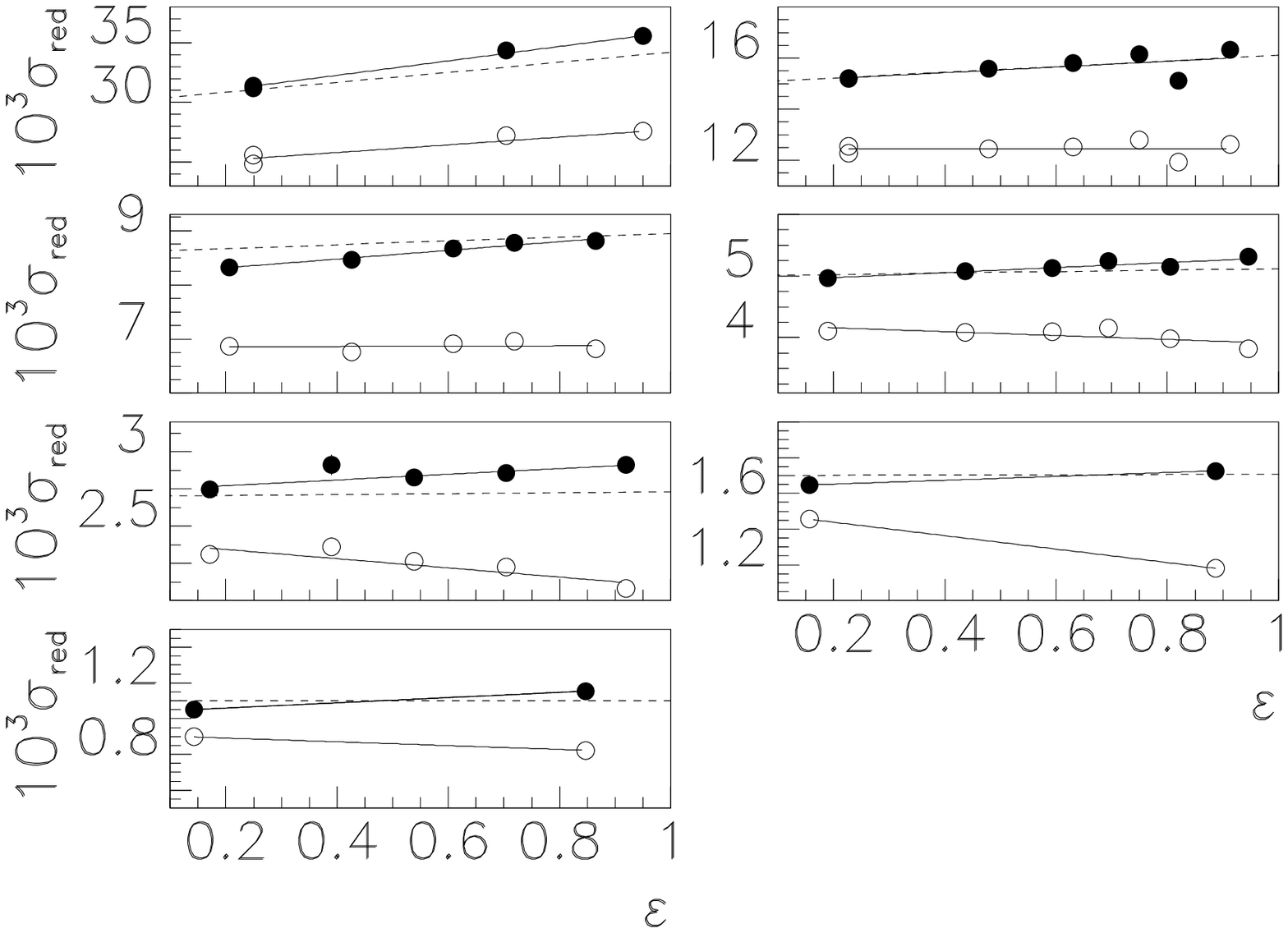}
\caption{\label{fig:sl1} Reduced cross section with (solid circle) and without (open circles) radiative corrections, for $Q^2$=1.75, 2.5, 3.25, 4, 5, 6, and 7   GeV$^2$. Data are from Ref. \protect\cite{An94}. Two-parameters linear fits are shown as solid lines. The dashed lines show the slopes suggested by the polarization data.}
\end{center}
\end{figure}

Therefore Eq. (\ref{eq:sred}) can be rewritten as:
$$
\sigma^{red}_{meas}(Q^2,\epsilon)= \sigma^{red}(Q^2,\epsilon)[1-\delta_R(Q^2,\epsilon)]\simeq G_M^2(Q^2) 
\left (\tau +\epsilon\displaystyle\frac{G_E^2(Q^2)}{G_M^2(Q^2)}
\right )[1-\epsilon\delta '(Q^2)]$$
\begin{equation} 
\simeq G_M^2(Q^2)\left[\tau+\epsilon\left (\displaystyle\frac{G_E^2(Q^2)}{G_M^2(Q^2)}-\tau\delta '(Q^2)\right )\right ],
\label{eq:sred2r}
\end{equation}
neglecting the term in $\epsilon^2$. At a fixed $Q^2$ (dropping the $Q^2$-dependence) we can see that the slope of  the measured cross section has also a linear behavior with $\epsilon$, with a slope $a$:
\begin{equation}
a={G_E^2}-\tau\delta'{G_M^2}.
\label{eq:sred3}
\end{equation}
This shows that the slope of the {\it measured} reduced cross section, as a function of $\epsilon$ (which is related to the electric FF squared), can vanish, and, for $\delta' > \displaystyle\frac{G_E^2}{\tau G_M^2}$ may even become negative. 
In Fig. \ref{fig:sl1} we show $\sigma_{red}$ as a function of $\epsilon$, with and without radiative corrections, taken from Ref. \cite{An94}. One can see that the slope is compatible with zero, for the uncorrected data, starting from $Q^2\le 2 $ GeV$^2$, and, then, becomes negative. 

The extraction of $G_E$ from the data requires a large precision in the calculation of radiative corrections, and in the procedure used to apply to the data. In particular at large $\epsilon$, an overestimate of the radiative corrections, can be source of a large change in the slope.

A similar study has been done from the other sets of data and leads to similar results.

\section{Conclusions}

From this analysis it appears that the available data on $ep$ elastic scattering do not show any evidence of deviation from the linearity of the Rosenbluth fit, and hence of the presence of the two photon contribution, when parametrized according to Eq. (\ref{eq:sfit}). 

The new generation of experiments performed at large $Q^2$ makes use of large acceptance detectors and requires important corrections of the raw data for acceptance, efficiency,  energy and angle calibrations \cite{Ch04}. Limits of the procedure and the approximations used for calculating radiative corrections were previously discussed in the literature \cite{Wa94}. 

Radiative corrections at the large $Q^2$ reached in the considered experiments may reach 30-40\%, whereas they are considered to be negligible in polarization eperiments (although one should stress that, in this case, no complete calculation is available). In particular they contain an important $\epsilon$ dependence which has a large influence on the slope of the reduced cross section, changing even its sign, as shown in this work. The slope is directly related to the extracted electric form factor. 

A careful study of the effect of radiative corrections to the experimental data seems necessary, due to their large size and large $\epsilon$ dependence,in particular for the large values of $\epsilon$, where the recent experiments have been performed. 

\section{Aknowledgments}

These results would not have been obtained without enlightning discussions with M. P. Rekalo.

Thanks are due to C. Perdrisat, V. Punjabi and L. Pentchev for useful remarks, and to J. Arrington and M. E. Christy for providing us with data in a tabulated form.

{}
\clearpage\newpage

\clearpage\newpage

\clearpage\newpage

\begin{thebibliography}{}

\bibitem{Ho62} 
R.~Hofstadter, F. Bumiller and M. Yearian,
 Rev.\ Mod. Phys. \ {\bf 30}, 482 (1958).

\bibitem{Shwinger}
J.~S.~Schwinger,
Phys.\ Rev.\  {\bf 76},  790(1949).

\bibitem{Wa94} 
R.~C.~Walker {\it et al.},
Phys.\ Rev.\ D {\bf 49}, 5671 (1994).

\bibitem{Ro50} M. N. Rosenbluth, Phys. Rev.  {\bf 79}, 615 (1950).

\bibitem{An94}
L.~Andivahis {\it et al.},
Phys.\ Rev.\ D {\bf 50}, 5491 (1994).
\bibitem{Ar75} 
  R.~G.~Arnold {\it et al.},
  Phys.\ Rev.\ Lett.\  {\bf 57}, 174 (1986);
  A.~F.~Sill {\it et al.},
  Phys.\ Rev.\ D {\bf 48}, 29 (1993).
\bibitem{Jo00}
M.~K.~Jones {\it et al.}  [Jefferson Lab Hall A Collaboration],
Phys.\ Rev.\ Lett.\  {\bf 84}, 1398 (2000).

\bibitem{Ga02}
O.~Gayou {\it et al.}  [Jefferson Lab Hall A Collaboration],
Phys.\ Rev.\ Lett.\  {\bf 88}, 092301 (2002).

\bibitem{Re68} A. Akhiezer and M. P. Rekalo, Dokl. Akad. Nauk USSR, {\bf 180},
1081 (1968); Sov. J. Part. Nucl. {\bf 4}, 277 (1974).

\bibitem{04108} Jefferson Lab E04-108 Proposal, 'Measurement of $G_{Ep}/G_{Mp}$ to
$Q^2$=9 GeV$^2$ via Recoil Polarization', (Spokepersons: E. Brash, M. Jones, C.F. Perdrisat, V. Punjabi).
\bibitem{Br03} 
E.~J.~Brash, A.~Kozlov, S.~Li and G.~M.~Huber,
Phys.\ Rev.\ C {\bf 65} 051001(R) (2002).

\bibitem{Ar04a}
J.~Arrington,
Phys.\ Rev.\ C {\bf 68}, 034325 (2003).

\bibitem{Ch04}
M.~E.~Christy {\it et al.}  [E94110 Collaboration],
Phys.\ Rev.\ C {\bf 70}, 015206 (2004).
\bibitem{Ar04}
I.~A.~Qattan {\it et al.},
arXiv:nucl-ex/0410010.


\bibitem{Mo69} L. W. Mo and Y. S. Tsai, Rev. Mod. Phys. {\bf 41}, 205 (1969).


\bibitem{Ma00} 
L.~C.~Maximon and J.~A.~Tjon,
Phys.\ Rev.\ C{\bf 62}, 054320 (2000);\\
A.~Afanasev, I.~Akushevich and N.~Merenkov,
Phys.\ Rev.\ D {\bf 64}, 113009 (2001);\\
A.~V.~Afanasev, I.~Akushevich, A.~Ilyichev and N.~P.~Merenkov,
Phys.\ Lett.\ B {\bf 514}, 269 (2001).

\bibitem{Gu73}  J. Gunion and L. Stodolsky,  Phys. Rev. Lett. {\bf 30}, 345
(1973);\\
 V. Franco,  Phys. Rev. D {\bf 8}, 826 (1973); \\
 V. N. Boitsov, L.A. Kondratyuk and V.B. Kopeliovich, Sov. J.
Nucl. Phys. {\bf 16}, 237 (1973); \\
  F. M. Lev, Sov. J. Nucl. Phys. {\bf 21}, 45 (1973).

\bibitem{We01}
S.~P.~Wells {\it et al.}  [SAMPLE collaboration],
Phys.\ Rev.\ C {\bf 63}, 064001 (2001).

\bibitem{Ma05}
  F.~E.~Maas {\it et al.},
  Phys.\ Rev.\ Lett.\  {\bf 94}, 082001 (2005).



\bibitem{Re99} M. P. Rekalo, E. Tomasi-Gustafsson and D. Prout, Phys. Rev.
{\bf  C60}, 042202(R) (1999).


\bibitem{Al99} L. C. Alexa {\it et al.}, Phys. Rev.  Lett. {\bf 82}, 1374 
(1999).
\bibitem{Ab99} D. Abbott {\it et al.}, Phys. Rev. Lett {\bf 82}, 1379 (1999).
\bibitem{Bl03} 
P.~G.~Blunden, W.~Melnitchouk and J.~A.~Tjon,
Phys.\ Rev.\ Lett.\  {\bf 91}, 142304 (2003).

\bibitem{Gu03} 
P.~A.~M.~Guichon and M.~Vanderhaeghen,
Phys.\ Rev.\ Lett.\  {\bf 91}, 142303 (2003).

\bibitem{Chen04}
Y.~C.~Chen, A.~Afanasev, S.~J.~Brodsky, C.~E.~Carlson and M.~Vanderhaeghen,
Phys.\ Rev.\ Lett.\  {\bf 93}, 122301 (2004).


\bibitem{Re1}
M.~P.~Rekalo and E.~Tomasi-Gustafsson,
Eur.  Phys. J. A. {\bf 22}, 331 (2004).


\bibitem{Re03t}
M.~P.~Rekalo and E.~Tomasi-Gustafsson,
Nucl.\ Phys.\ A {\bf 740}, 271 (2004).

\bibitem{Re04a}
M.~P.~Rekalo and E.~Tomasi-Gustafsson,
Nucl.\ Phys.\ A {\bf 742}, 322 (2004).

\end{thebibliography}
\end{document}